\documentclass[12pt,onecolumn]{emulateapj}
\usepackage{epsf}
\usepackage{epsfig}
\usepackage{ulem}
\newcommand{\kms}{km s$^{-1}$}

\def\wisk#1{\ifmmode{#1}\else{$#1$}\fi}

\begin{document}

\title{Cosmic Microwave Background filters and the Dark Flow measurement}

\author{F. Atrio-Barandela\altaffilmark{1},
A. Kashlinsky\altaffilmark{2}
H. Ebeling\altaffilmark{3},
D. Kocevski\altaffilmark{4}}
\altaffiltext{1}{F{\'\i}sica Te\'orica, Universidad de Salamanca,
37008 Salamanca, Spain; atrio@usal.es}
\altaffiltext{2}{SSAI and Observational Cosmology Laboratory, Code
665, Goddard Space Flight Center, Greenbelt MD 20771; Alexander.Kashlinsky@nasa.gov}
\altaffiltext{3}{Institute for Astronomy, University of Hawaii, 2680
Woodlawn Drive, Honolulu, HI 96822; ebeling@ifa.hawaii.edu}
\altaffiltext{4}{Department of Physics, University of California at Davis, 
1 Shields Avenue, Davis, CA 95616, USA; kocevski@physics.ucdavis.edu}

\begin{abstract}

Recent measurements of large-scale peculiar velocities from the cumulative 
kinematic Sunyaev-Zel'dovich effect using WMAP data and X-ray selected
clusters from ROSAT have identified a bulk flow 
of galaxy clusters at $\sim 600-1,000$ km s$^{-1}$ on scales of $\sim0.5-1$ Gpc, 
roughly aligned with the all-sky Cosmic Microwave Background (CMB) dipole. 
The flow is inferred from the detection of a residual dipole generated 
by CMB fluctuations exclusively in the direction of galaxy clusters, and measured
within apertures containing zero monopole. Consistent with this interpretation, 
the amplitude of the dipole correlates with the X-ray luminosity of the clusters. 
To enable this measurement, the CMB data need to be filtered to 
remove the primary CMB, thereby increasing the data's signal-to-noise ratio.
Filtering cannot imprint a signal with the mentioned properties at cluster positions; 
however, an inadequately designed filter can greatly suppress such a signal.
We show here that recent studies that failed to detect 
a large-scale flow with various filters
indeed adopted flawed implementations; when correctly implemented, these alternative 
filters lead to results that are in fact consistent with the Dark Flow signal.
The discrepancies can be traced to the likely presence of residual dipoles caused by the
thermal Sunyaev-Zel'dovich effect, to assumptions about cluster
profiles incompatible with the data as well as failure to compute dipoles at
the zero monopole aperture. PLANCK maps, with their large frequency coverage and a 217 
GHz channel, will be instrumental to probe bulk flows, to 
remove spurious dipole signals and to help identify filtering 
schemes appropriate for this measurement.
\end{abstract}


\section{Introduction.}

Peculiar velocities are deviations from the uniform expansion of
the Universe. In the standard gravitational-instability paradigm, they are
generated by inhomogeneities in the matter distribution. Measurements of peculiar 
velocities using galaxies rely on distance indicators to subtract the Hubble 
expansion, which limits the volume probed to $\sim 100h^{-1}$Mpc 
(see  Strauss \& Willick 1995; Kashlinsky, Atrio-Barandela \& Ebeling 2012 for reviews). 
As an alternative, Kashlinsky \& Atrio-Barandela (2000, hereafter KAB) proposed a method
to probe the large-scale velocity field via the dipole induced by anisotropies in 
the CMB temperature that are caused by the kinematic Sunyaev-Zel'dovich (SZ; 
Sunyaev \& Zel'dovich 1970, 1972) effect in the direction of clusters of galaxies.
The SZ has two components: a thermal one (TSZ), caused by the thermal motions of 
electrons in the potential wells of clusters, and a kinematic one (KSZ), 
caused by the motion of the cluster as a whole
(kinematic SZ, KSZ; see e.g. the review by Birkinshaw 1999). 
Once it is imprinted on the CMB, the SZ distortion remains fixed. It does not 
depend on distance, making it a very useful tool to detect clusters at high redshift 
(Planck Collaboration, Planck Early results XXVI, 2011).  

The KSZ effect measures motion directly with respect to the CMB frame and 
does not require subtraction of the Hubble expansion.  The KAB method 
evaluates the dipole from an all-sky CMB temperature map at the locations 
of known X-ray luminous clusters. Since the main contaminant in
foreground clean maps is the primary CMB 
signal, the data need to be filtered to allow the peculiar velocities
of clusters to be measured (Haehnelt \& Tegmark 1996);
KAB proposed to use a Wiener-type filter for this purpose.  
In Kashlinsky et al (2008, hereafter KABKE, 2009) we showed that,
within cluster apertures containing zero CMB monopole, a residual CMB dipole 
exists which is aligned with the CMB dipole. The amplitude of this 
residual dipole scales with 
cluster X-ray luminosity (Kashlinsky et al 2010; Kashlinsky, Atrio-Barandela 
\& Ebeling 2011) proving that the signal is indeed associated with clusters.
We have interpreted this dipole as the signature of a bulk flow. If the flow extends 
to the cosmological horizon, part of the all-sky CMB dipole could be viewed 
as being of primordial, rather than kinematic, origin. As of this writing, 
no alternative interpretations of the signal, which would have been 
of considerable interest, have been offered in peer-reviewed literature.

Our results have been challenged by other authors.
Keisler (2009), as well as the contemporaneous study by 
Kashlinsky et al (2010), noted that the residuals for each 
of the maps for the eight WMAP DAs  (Differential Assemblies) 
in the KABKE measurement are correlated, which
increases the errors quoted by KABKE, but by no more
than $\sqrt{N_{\rm DA} }=2.8$, where $N_{\rm DA}=8$ is the number of DAs
of the WMAP dataset. Keisler then claimed that the 
detection of a large-scale flow was in fact not 
statistically significant, his error corrections exceeding the 
$\sqrt{N_{\rm DA}}$ factor. Atrio-Barandela et al (2010) 
pointed out that Keisler (2009)
had failed to subtract the dipole of the filtered maps outside
the Galactic mask, thereby overestimating his error bars. 
In addition, we also proved
that the filter devised by KAB removes the primary CMB down to the
fundamental limit  of cosmic variance (Atrio-Barandela et al 2010). 
It was this highly 
efficient filtering scheme that allowed a detection of the flow at 3.5--4$\sigma$
confidence level, depending on the cluster sample used. 
All data needed to verify the measurement have been made publicly
available\footnote{\url{http://www.kashlinsky.info/bulkflows/data\_public}}
for examination by the interested reader. 

Osborne et al (2011, OMCP) studied the effect of different filters on 
bulk-flow measurements from WMAP data. In their Appendix B.2 OMCP reproduced 
our results at the $2.9\sigma$ confidence level but failed to notice that 
this in itself invalidates the statistical uncertainties
claimed by Keisler (2009). The lowered significance (compared to our 3.5--4 $\sigma$), 
is due to differences in the catalogs used (Kashlinsky et al  2011) 
and to the relatively small number of simulations ($\sim 100$) carried 
out by OMCP to compute the errors. OMCP also implemented alternative, 
``matched'' filters, with which they  found no evidence of bulk flows in WMAP data. 
Those matched filters, however, were 
designed more to detect radio point sources than to remove the primary CMB 
anisotropies. As a result, they were insensitive to bulk motions to the extent that 
the KSZ signal is weaker in the filtered maps than in the original 
data.  As can be seen in OMCP's Fig.~13, a sample of $\sim 730$ clusters 
moving at velocities  of $\sim 4,000-10,000$ km s$^{-1}$ is required to produce 
a detectable signal in the filtered maps. At such extreme velocities, the 
motion of individual clusters should be detectable directly in the 
original CMB maps without any need for filtering. 

In another study, Mody \& Hajian (2012, MH), using a 
different Wiener filter, concluded that the measured bulk-flow amplitude
was ``consistent with the $\Lambda$CDM prediction". 
This statement (made prominently in the MH abstract) is largely 
meaningless though, since the MH error bars of  1,400--2,600 \kms\ 
per velocity component render their methodology unable to discriminate 
between the concordance $\Lambda$CDM prediction and a bulk flow of amplitude 
600--1,000 km s$^{-1}$, as measured by KABKE.

In spite of the aforementioned problems with these studies, both the OMCP and 
MH results have been cited repeatedly as evidence against 
the existence of a Dark Flow (DF).  In this paper, we examine in more 
detail why the filters used by OMCP and MH failed to detect a 
statistically significant dipole signal. We stress that our focus is on 
the dipole signal, {\it irrespective of its interpretation}. Specifically, we
aim to answer the question of why the dipole signal was not detected with 
the other filtering schemes. Do they reduce the S/N 
to levels that render the dipole statistically insignificant? Do they 
redistribute the dipole signal to lie outside 
the aperture ranges probed there? Equally important, none of the
proposed new filters have been demonstrated to probe the same flow
than the KAB filter, so the discrepancies between filters could be
due to variations of the flow. In this article 
we will show that the OMCP and MH filters are not as efficient
as KAB, but that, when correctly implemented, either filter recovers 
a signal that is in fact compatible with the DF identified by us.
In Sec.~\ref{sec:filters} we discuss the properties 
of filters, their effect on the data, and summarize previous results.  
In Sec.~\ref{sec:sec3} we compute the CMB dipole at the locations of clusters in maps
processed with the different filters and discuss their relative
merits and their consistency, in Sec.~\ref{sec:summary} we summarize our
results and present our conclusions.

\section{KAB method and CMB filtering.}
\label{sec:filters}

At the positions of clusters, the microwave temperature anisotropies 
have four main components: the primary CMB, the TSZ and KSZ signal, and 
instrument noise. In foreground clean maps one needs to expect some
level of foreground residuals, but these residuals will not correlate with clusters in the
sky and outside the Galactic plane are negligible compared with the cosmological signal.
The KSZ anisotropy is given by $\Delta T_{\rm KSZ}= - T_0\tau (V_{\rm Bulk}/c)$, where
$\tau$ is the projected electron density along the line of sight,
$V_{\rm Bulk}$ the bulk-flow velocity of our cluster sample, $c$ the speed of light, 
and $T_0$ the CMB blackbody temperature. The bulk flow motion of all
clusters in a given sample will generate a CMB dipole at the cluster
location that  we estimated it to be (KAB)
\begin{equation}
a_{1m}=1\mu {\rm K} \left(\frac{V_{\rm bulk}}{300 \;{\rm km\; s^{-1}}}\right)
\pm 3\mu {\rm K}\left(\frac{N_{\rm cl}}{\mbox{1,000}}\right)^{1/2}
\pm 0.6\mu {\rm K}\left(\frac{N_{\rm pix}}{\mbox{10,000}}\right)^{1/2}\pm
0.2\mu {\rm K}\left(\frac{N_{\rm cl}}{\mbox{1,000}}\right)^{1/2}
\label{eq:dipole}
\end{equation}
where the terms on the right-hand side describe the contributions 
from the KSZ, the cosmological CMB, noise, and the TSZ, respectively. 
In this expression, 
$N_{\rm cl}$ clusters that occupy a solid angle of $N_{\rm pix}$ pixels in 
the sky. Instrument noise (the third term) can be further reduced as 
$N_{\rm DA}^{-1/2}$ by combining $N_{\rm DA}$ differential assemblies 
operating at different frequency bands. In addition, the fourth term, 
reflecting the contribution from the TSZ effect, can be isolated and 
eliminated by using its distinctive frequency dependence. 
Nevertheless, the TSZ dipole due to the inhomogeneous distribution
of clusters in the sky is bounded above by the TSZ monopole. In
Kashlinsky et al (2008, 2009) we proposed to measure dipoles at zero monopole aperture
to limit the TSZ dipole contribution even without frequency information.

The largest source of errors in eq.~(\ref{eq:dipole}) is 
the sample-variance uncertainty caused by the small number of 
pixels at which the dipole is evaluated. To determine peculiar velocities 
it is vital to increase the signal-to-noise ratio (S/N) for these pixels 
by reducing the contribution from the primary CMB anisotropy. However, 
unlike for the TSZ component, frequency information can not be 
used to discriminate between the CMB and KSZ contributions to the overall 
dipole. To address this problem, KAB proposed to use the statistical properties 
of the primary CMB to design a filter that effectively removes its 
contribution while preserving the KSZ component, thereby increasing the 
S/N of the proposed measurement. 

Reflecting the spatial isotropy of the CMB temperature anisotropies, 
filters are spherically symmetric; in $\ell$-space, the filter 
depends on $\ell$ but not on $m$. If $\Delta T=\sum_{\ell m} a_{\ell m} Y_{\ell m}$
and $F(\theta)=(1/4\pi)\sum_\ell (2\ell+1)F_\ell^2 P_\ell(\cos\theta)$ are
the expansions of the temperature map in spherical harmonics and of the filter 
in Legendre polynomials, respectively, then the filtered map is given by
\begin{equation}
F\star\Delta T=\sum_{\ell m} F_\ell a_{\ell m} Y_{\ell m}\,,
\label{eq:filter}
\end{equation}
where $\star$ denotes convolution. In the following sections 
we will discuss the distribution of a KSZ dipole in $\ell$-space,
followed by a brief description of the filters under investigation here.

\subsection{The KSZ signal.}

\begin{figure}
\plotone{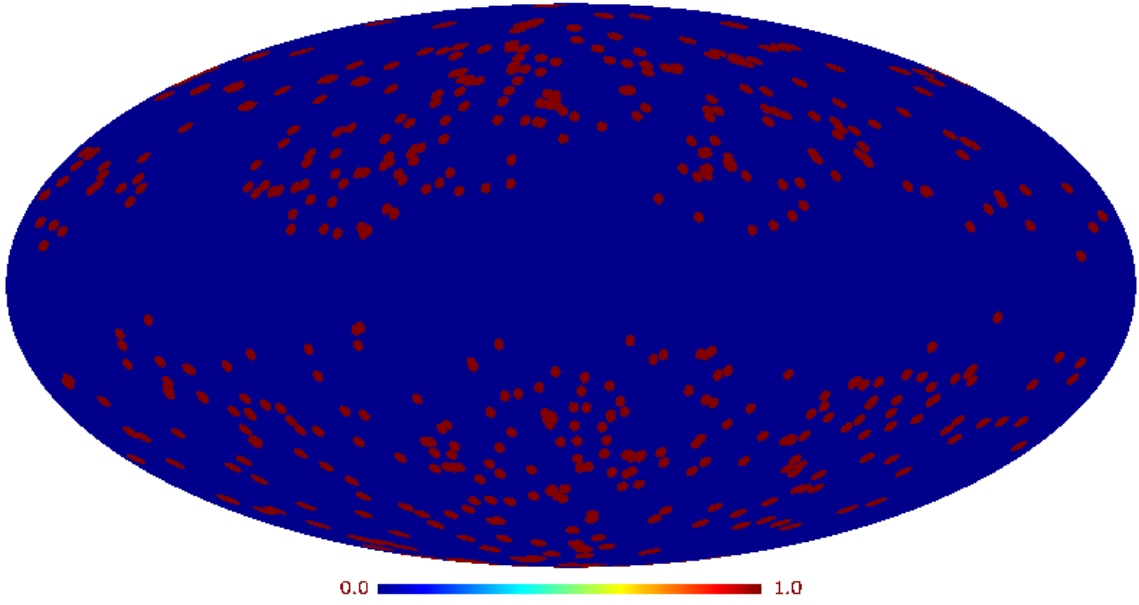}
\caption{\small 
Distribution on the
sky of the 506 clusters of our sample in Healpix format. Clusters are plotted
as discs of radius $1.5^\circ$ for easier visualization.
}
\label{fig:sample}
\end{figure}

Our bulk-flow studies were based on an all-sky cluster sample created by 
combining the ROSAT-ESO Flux Limited X-ray catalog (REFLEX) 
(B\"ohringer et al. 2004) in the southern hemisphere, the extended 
Brightest Cluster Sample (eBCS) (Ebeling et al. 1998, 2000) in 
the north, and the Clusters in the Zone of Avoidance (CIZA) 
(Ebeling et al. 2002; Kocevski et al. 2007) sample along the 
Galactic plane. Our full catalog contains $\sim 1500$ entries
including clusters and galaxy groups. To illustrate the performance
of the different filters, we select a subset of 506 clusters
located outside WMAP's KQ75 galactic and point source mask.
These clusters are located at redshifts $z\le 0.16$ and
have X-ray luminosities $L_X[0.1-2.4\; {\rm keV}]\ge 0.5\times 10^{44}erg\,s^{-1}$.
For the purposes of this paper we restrict our analysis to clusters
for which the parameters of a $\beta$ profile (Cavaliere \& Fusco-Femiano 1976) 
have been measured by fitting to ROSAT X-ray data. This sample is smaller 
than the one used in Kashlinsky et al (2010) and our final statistical
significance will be degraded with respect to our earlier results.
We divided our set into three independent 
samples, selected according to luminosity. We define three
intervals: $L_X\in([0.5,1],[1.,2],[>2])\times 10^{44}$ erg s$^{-1}$.
The distribution of our clusters on the sky is shown in 
Fig.~\ref{fig:sample} using a Healpix projection of $N_{side}=512$
(Gorski et al., 2005). All clusters are drawn as discs of radius $1.5^0$.

\begin{figure}
\plotone{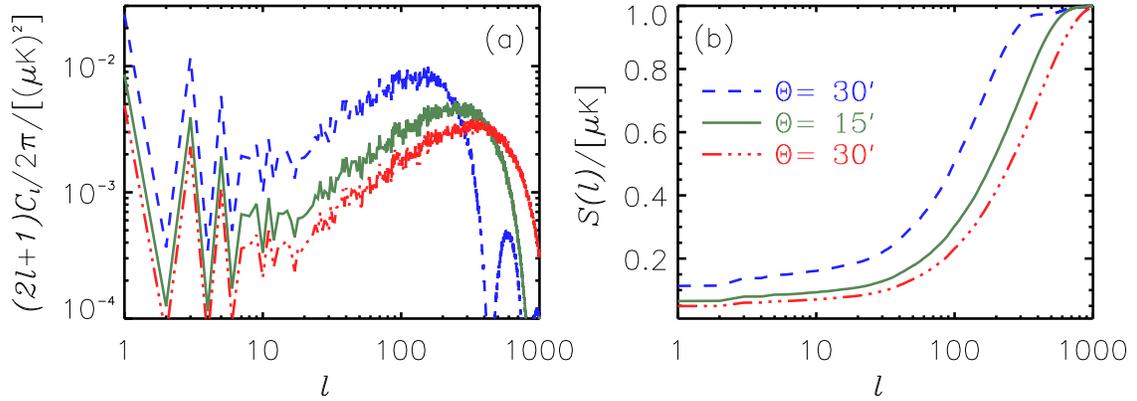}
\caption{\small (a) Power spectrum of a KSZ dipole generated by 506 identical
clusters of different sizes: 30\arcmin\ (dashed blue line), 15\arcmin\ (solid green)
and 10\arcmin\ (dot-dashed red), moving in the direction of the CMB dipole
with an amplitude of $D=1\mu$K. (b) Integrated signal (eq.~\ref{eq:signal})
for the same three angular sizes; lines follow the same convention than in (a).
}
\label{fig:signal}
\end{figure}

The set of clusters in the final sample occupies less than 1\% of the sky. 
In such a  small solid angle, spherical harmonics are not orthogonal.
Before discussing how filters operate, we need to discuss how 
the dipole power at $\ell=1$, that is evaluated only at the 
cluster positions, is redistributed among all multipoles. 
If $a^{\rm KSZ}_{\ell,m}$
are the spherical harmonic transforms of the KSZ component
of the KSZ dipole evaluated at $N_{\rm cl}$
locations in the sky then the power at 
each multipole $\ell$ is $(2\ell+1)C_\ell^{\rm KSZ}=
\sum_{m}|a^{\rm KSZ}_{\ell,m}|^2$.
We define $S(\ell)$ as the contribution 
of all multipoles $i\le\ell$ to the total dipole in the map as: 
\begin{equation}
S^2(\ell)=\frac{1}{4\pi}\sum_{i=2}^{\ell}(2\ell+1)C_\ell^{\rm KSZ}.
\label{eq:signal}
\end{equation}
To predict $C_\ell^{KSZ}$ and $S(\ell)$ we need
the electron-density profile of all clusters. 
Suitable data to derive these profiles accurately are not yet available.
To compute the dipole power we will assume that all clusters subtend 
the same solid angle and move coherently in 
the direction $(l,b)=(270^\circ,30^\circ)$. For simplicity 
we assign a dipole anisotropy $D=-1\mu$K$\cos\theta$ to all pixels of each cluster,
where $\theta$ is the angle between the line of sight of the
cluster and the apex of the motion.
Figs.~\ref{fig:signal}a and \ref{fig:signal}b depict the resulting power and 
the integrated dipole signal $S(\ell)$, respectively, for identical 
clusters of angular radius 10\arcmin, 15\arcmin\ and 30\arcmin. 
Fig.~\ref{fig:signal}a shows that the lowest multipoles have the largest
amplitude, and that power is transferred preferentially to odd multipoles.
Fig.~\ref{fig:signal}b shows that
less than 10\% of the signal remains at $\ell=1$; the bulk of the signal is 
contained in the range $1<\ell<300$ even for unresolved clusters. 

The exact distribution of the KSZ dipole in $\ell$-space will
depend on the (unknown) profile of clusters and their extent. Since clusters
have a radius of 10\arcmin-30\arcmin, the effect of the cluster profile will
be important for multipoles $\ell>200-600$. Therefore, our conclusion 
that more than 50\% of the KSZ dipole signal lies below those scales
is not affected by taking a constant profile. 
The actual signal will be a combination of the contributions from resolved 
(15\arcmin--30\arcmin) and unresolved (10\arcmin) clusters, with clusters 
of different mass contributing at different relative weights. 

\subsection{The KAB filter.}

The KAB filter was specifically designed to remove the primary CMB of the 
measured cosmological model. The filter is of the Wiener type and was
constructed to  minimize the contribution from the cosmological signal in the 
filtered maps in the presence of noise, ${\cal N}$, i.e.\ it minimizes 
$\langle(\Delta T-{\cal N})^2\rangle$ (Kashlinsky et al 2009). 
In $\ell$-space the filter is given by
\begin{equation}
F^{\rm KAB}_\ell=\frac{C_\ell^{\rm sky}-C_\ell^{\rm \Lambda CDM}B_\ell^2}
{C_\ell^{\rm sky}}\,,
\label{eq:kab}
\end{equation}
where $C_\ell^{\rm sky}$ is the actual realization of the radiation power
spectrum in the sky that includes noise, TSZ, KSZ, foreground residuals,
and primary CMB; $C_\ell^{\rm \Lambda CDM}$ is the power spectrum
of the $\Lambda$CDM model that best fits the data, and $B_\ell$ is the antenna
beam for a given DA. Since the quadrupole and octupole are
aligned with the dipole, the KAB filter is set to zero for $\ell\le 3$
to avoid any cross-talk from the corresponding large angular scales that could mimic
a dipole. In Fig.~\ref{fig:filters}a we represent the KAB filter for the W1 DA. The filter
has structure at all multipoles and is therefore sensitive to KSZ power
across the full multipole range.

\subsection{OMCP filters.}\label{subsec:omcp}

OMCP used Matched Filters (MF), i.e., filters that are suited to distinguish 
point sources from a background signal, provided the profile of the source 
is known. In their analysis, OMCP assumed that all clusters are unresolved and
approximated their profiles by the WMAP beam function. In $\ell$-space this means
\begin{equation}
F_\ell^{\rm OMCP}=
\frac{B_\ell}{(C_\ell^{\rm \Lambda CDM}+C_\ell^{\rm TSZ})B_\ell^2+N_\ell}\,,
\label{eq:omcp}
\end{equation}
where $C_\ell^{\rm TSZ}$ and $N_\ell$ are the power spectrum
of the TSZ effect and of the white and homogeneous instrumental noise, 
respectively. $B_\ell$ is the antenna beam. In addition, OMCP considered clusters
to be isothermal, in which case the TSZ and KSZ profiles are proportional to
each other. By measuring the TSZ component in the original and in the filtered data,
the fraction of the KSZ lost due to filtering could be computed exactly.
However, since clusters are \textit{not}\/ isothermal
(as demonstrated, e.g., by Atrio-Barandela et al.\ 2008), their electron-density 
and pressure profiles are in fact different, causing filters that disregard
this fact to potentially suppress the KSZ signal relative to the TSZ contribution.
In Fig.~\ref{fig:filters}b we plot the OMCP filters for eight WMAP DAs,
corresponding to the Q (dot-dashed green line), 
V (dashed blue) and W (solid red) bands. These filters
are effectively band-pass filters, removing all power below $\ell\sim 200-300$ and 
above $\ell\sim 800-1000$. 

OMCP also considered a modified MF, named Unbiased Multifrequency
Matched Filter (UF), designed to remove the TSZ contribution. Since this second
filter was found to be even less sensitive to bulk flows than MF, we will not consider
it further.  Our approach to reducing any TSZ contamination will be 
the same for all filters, namely to compute dipoles at the zero-monopole aperture.

\subsection{The Mody-Hajian filter.}\label{subsec:mh}

MH used a Wiener filter 
designed to remove the TSZ and instrument noise
contributions together with  the intrinsic CMB, leaving only 
the KSZ component in the data. Their filter is defined by
\begin{equation}
F^{\rm MH}_\ell=\frac{C_\ell^{\rm KSZ}B_\ell^2}
{[C_\ell^{\rm \Lambda CDM}+{C}_\ell^{\rm TSZ}]B_\ell^2+N_\ell}\,,
\label{eq:mh}
\end{equation}
where $C_\ell^{\rm KSZ}$ is the power spectrum of the KSZ contribution,
and the other terms have the same meaning as in eqs.~(\ref{eq:kab}, \ref{eq:omcp}).
This filter neglects the inhomogeneities in the instrumental noise
and assumes that $C_\ell^{\rm KSZ}$, the quantity to be
measured, is known. To construct their filter,
MH need to 1) specify the cluster distribution on
the sky, 2) assume  an electron density profile for all clusters, and
3) fix the cluster extent, all quantities that have yet to be measured from the data. 
In Fig.~\ref{fig:filters}c we plot the MH filter for assumed cluster 
extents of 10\arcmin\ and 15\arcmin; both are convolved with the 
beam of the W1 DA. The filter is normalized to a KSZ  dipole in 
the Z-direction with an amplitude of $10\mu$K;
convolution with the beam reduces the amplitude to $\sim 5\mu$K.
The resulting two curves can be thought of as two extreme cases
of what the actual MH filter might be. Taking into account that
very few clusters are resolved in the WMAP W band, we
will in the subsequent discussion only consider the MH filter constructed 
assuming an angular radius of 10\arcmin.

\subsection{Cluster apertures on the filtered data.}

Filtering not only removes the intrinsic CMB component, it also modifies
the KSZ signal, affecting the cluster extent and its electron
density profile.  To evaluate 
the monopole and dipole  at $N_{cl}$ cluster positions on the sky we first need 
to discuss what would be the radial extent of each cluster in the filtered data. 
In Kashlinsky et al (2008) we considered that clusters extended 4--6 times
the radial extent of the region emitting 99\% the total X-ray flux. Since 
Kashlinsky et al. (2010) we quote instead
the dipole evaluated on discs of $\sim 25$\arcmin\ aperture radius, the same size
for all clusters. In both cases, the monopole
was close to zero and the measured dipole did not differ significantly.
{\it Evaluating dipoles at the zero monopole aperture is crucial to guarantee 
that the measured dipole has no significant contribution from the TSZ effect}: 
since the monopole is dominated by the TSZ contribution, the amplitude of the 
former is an upper limit to any dipole due 
to the inhomogeneous distribution of clusters on the sky.

\begin{figure}
\plotone{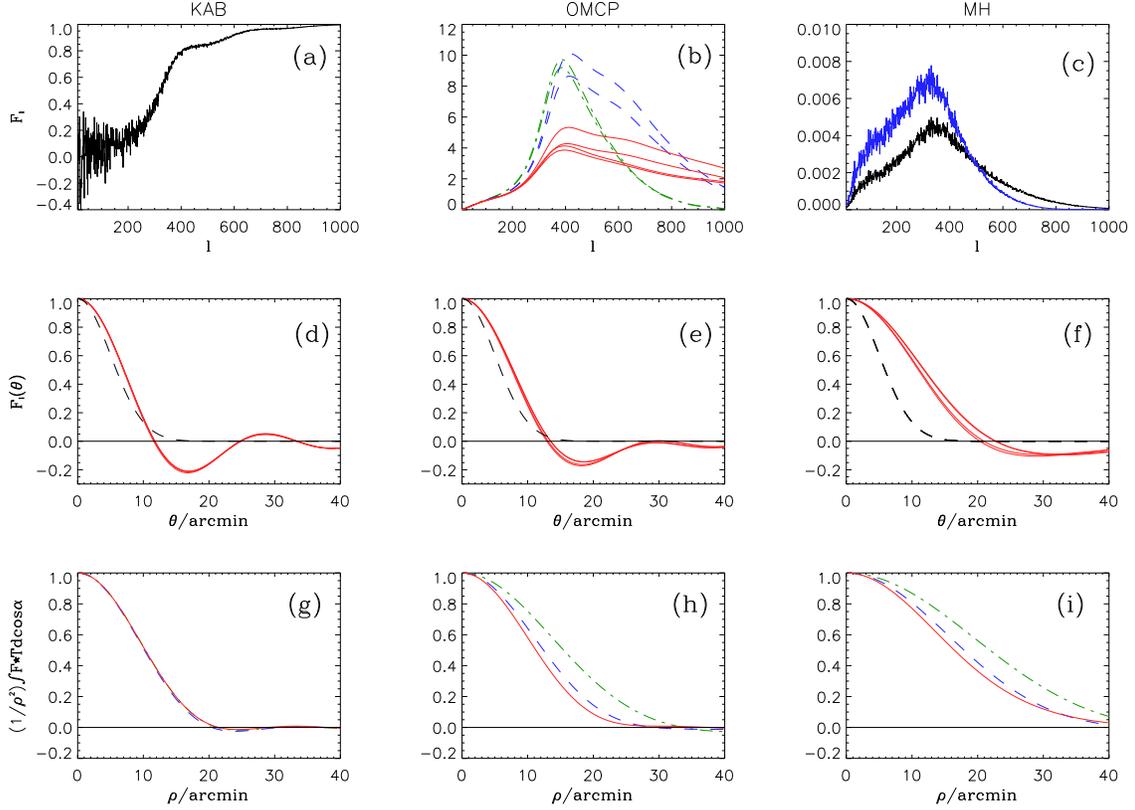}
\caption{\small Filters in (top to bottom) $\ell$ space (a-c), 
real space (d-f), and signal integrated
over a disc of angular radius $\rho$ (g-i) for 
(left to right) the three different 
filters defined by KAB, OMCP, and MH 
discussed in Sec.~\ref{sec:filters}. (a)
KAB filter for the WMAP W1 DA. (b) OMCP filters
for the Q (dot-dashed green line), V (dashed blue line) and 
W (solid red) channels. (c) MH filters, computed assuming 
that all clusters have the same angular size of either 10\arcmin\ (black line) or 
15\arcmin\ (blue line). (d--f) The respective W-band filters in real space (red,
solid line). The dashed line corresponds to a gaussian antenna with the resolution
of the W-band.
(g--i) Filter averaged over an aperture of size $\rho$ as a function
of aperture size. The lines follow the same convention as in (b).
}
\label{fig:filters}
\end{figure}

The cluster profile, $T(\theta)=(1/4\pi)\sum (2\ell+1) T_\ell P_\ell(\cos\theta)$,  
when convolved with the antenna
beam $B_\ell$ and the filter $F_\ell$, becomes:
\begin{equation}
(F\star T)(\alpha)=A\sum (2\ell+1)F_\ell T_\ell B_\ell P_\ell(\cos\alpha)\,,
\label{eq:profile}
\end{equation}
where $A$ is a normalization constant.
In Figs.~\ref{fig:filters}d--f we plot the filter profile in real space
of the W-band (solid red lines). 
This would be the cluster profile in the filtered map if 
it were a Dirac $\delta$-function, i.e., if the cluster was unresolved.
For comparison, the dashed line
represents Gaussian beam with the resolution of the W band.

To assess the impact of the aperture choice, we compute the measured 
cluster profile from eq.~(\ref{eq:profile}), averaged 
over a disc of angular radius $\rho$, as it was done in the data
\begin{equation}
\langle F\star T\rangle_\rho=\frac{1}{\rho^2}\int_{0}^\rho 
[(F\star T)(\alpha)] d\cos\alpha\,. \label{eq:filter_average}
\end{equation}
In Fig.~\ref{fig:filters}g--h we plot the above average vs the
angular radius $\rho$ of the disc for the KAB, OMCP, and MH filters, respectively. 
To avoid overcrowding the plots, we only show the Q1, V1 and W1 DAs. For the KAB 
filter the signal vanishes around 20\arcmin,
while it extends to 30\arcmin\ for the OMCP filters, and  well beyond
40\arcmin\ for the MH filter. These radial extents correspond to unresolved clusters;
for real sources the signal will spread to apertures of even {\it larger} radii. 
Empirically, we found that the zero monopole aperture was around 25\arcmin\ for the
KAB filter, whereas slightly above 20\arcmin\ would correspond to all clusters
being unresolved by the antenna beam. The zero monopole defines
a physically well motivated aperture over which to compute the dipole.
\textit{This point was missed by OMCP and MH,} who never have addressed this issue.

\subsection{Summary of previous results.}

The filters discussed in this section are spherically symmetric.
They cannot imprint a dipole at zero monopole exclusively at cluster locations 
on a filtered map but, if not carefully designed and implemented, they
can erase it to below the statistically significant level. 
OMCP implemented two matched filters, named 
MF, described in Sec.~\ref{subsec:omcp}, and UF.
They measured the dipole at the position of 736 clusters of galaxies
and found no significant KSZ signal. OMCP tested the performance 
of their filters using simulated maps and showed that 
their filters begin to recover the input velocity  only at speeds 
$V_{\rm bulk}\simeq 4,000$ km s$^{-1}$ for the MF and $V_{\rm bulk}\simeq 10^4$ 
km s$^{-1}$ for the UF. The average mean optical 
depth of the clusters in their sample is quoted as  
$\langle \tau\rangle \simeq 4.9\times 10^{-3}$ (Sec. 4.5 of OMCP), 
and the resulting temperature anisotropy would be $\Delta T=183\mu {\rm K}(\tau/0.005)
(V_{\rm bulk}/4000\;{\rm km\;s^{-1}})$. Velocities approaching or exceeding 
4,000 \kms\ thus induce anisotropies that would be measurable in the original 
data and for each individual cluster, without any need for filtering. 

The root cause of this problem is that OMCP's implementation of their 
filters did not increase the S/N but strongly reduced it, 
to the extent of rendering their filters unable to detect a CMB 
dipole corresponding to a bulk-flow velocity of $\sim 600-1000$ km s$^{-1}$. 
The poor performance of their filters is also due to the fact that OMCP
effectively measured the dipole only at the central cluster pixels 
because their filters were optimized to reconstruct
the source amplitude if the source was centered on that pixel
(Mak et al 2011). Since clusters in the filtered
maps reach a zero monopole aperture at $\sim$30-40\degr
(see Fig.~\ref{fig:filters}d--i) the implementation
by these authors does not use all the information available
in the filtered map. Also, OMCP assumed 
that the KSZ and TSZ cluster profiles are identical and unresolved. However, any 
filter that efficiently removes the TSZ signal will also remove a substantial 
part of the KSZ signal. This is the reason why the better designed UF requires
three times larger velocities than the MF to detect the average 
motion of $\sim 700$ clusters.

MH, by contrast, used a different approach. Instead of  computing the 
dipole at the locations of clusters, MH  fit dipole templates to their filtered 
maps. This approach is flawed in two important respects: (a) MH require 
that the total number of clusters, their profiles, and extents be known 
in advance in order to construct their filter, and (b)
by matching the same cluster template to the filtered data
MH implicitly assume that all clusters have the same 
profile and extent in the sky as in the filtered CMB data. 
Hence, changing the profile, number, or extent of clusters changes the
filter (and the filtered map) and the final results.
As both OMCP and MH restricted their analyses
to apertures smaller than the ones used by KABKE, the dipole measured by 
us could have remained undetected in the residual CMB and noise of their filtered maps.

\section{Measuring Bulk Flows in WMAP data with the KAB method.}\label{sec:sec3}

To measure the dipole at the locations of clusters in CMB maps 
filtered following KAB, OMCP, and MH, our pipeline proceeds as follows:
\begin{enumerate}
\item{} For each DA, the data are multiplied by the extended mask KQ75 for 
the WMAP7 data release. Then, monopole, dipole, and quadrupole are subtracted 
from the regions outside the mask. Next, we compute the multipole expansion 
coefficients $a_{\ell m}$. The power lost due to masking is corrected for by 
multiplying each multipole by $f_{\rm sky}^{-1/2}$, where  $f_{\rm sky}$ 
is the fraction of the sky outside the mask.
\item{} The $a_{\ell m}$ coefficients are multiplied by the filter $F_\ell$ before 
transforming them back into real space to create the filtered map. 
\item{} The monopole and dipole outside the mask are removed from
the filtered maps. 
\item{} Dipoles are computed at the cluster positions using the same 
aperture of radius $\rho$ for all clusters. We repeat the measurement
for different $\rho$ and, as indicated in Sec.~2.5, we select the 
dipole measured at the {\it zero monopole} aperture.
\end{enumerate}

To translate the measured dipole amplitude from temperature units to velocity units,
we need to compute how the mean cluster opacity changes with filtering.
Owing to the different functional form of the three filters, 
they will affect the KSZ power differently.  Since $\tau = \tau(\theta)$,
to compute the average $\langle\tau\rangle$  over our cluster sample we
need the electron density profile and of the cluster radial extent 
of all clusters, information that is not presently available.
In the KAB filter, this uncertainty translates into a calibration
uncertainty but does not affect the amplitude of the dipole in
$\mu$K nor its direction. As discussed, this uncertainty is built 
into the definition of the OMCP and MH filters. In order to facilitate
the comparison of the results obtained with the different
filtering schemes, we will use the results presented in Figs.~\ref{fig:filters}d--f 
to normalize the filtered maps as described below.

\subsection{Filter normalization and filtered maps.}

The filters defined in Sec.~2 and depicted in Fig.~\ref{fig:filters}a-c
do not only differ in shape but also in normalization.
To normalize a map we need to derive a coefficient $R^F$ that 
measures how much the dipole is diluted by the filter. 
If $\tau(\theta)$ and $F\star\tau(\theta)$
are the cluster optical depth in the unfiltered and filtered maps,
then the normalization coefficient would be 
$R_F=\langle \tau\rangle/\langle F\star\tau\rangle$,
where the average is taken over the solid angle subtended by
the clusters in unfiltered and filtered data. This parameter 
is not simple to compute since both profiles and angular extents are unknown.
Instead, an approximated normalization coefficient can be computed by 
taking into account that relaxed clusters are close to isothermal 
in their central regions (Pratt et al.\ 2011) and electron pressure and 
electron density are proportional to each other. Hence, in
those regions \textit{filtering would dilute the TSZ and KSZ signal 
in the same proportion.} Then,
$R^F\approx\langle \Delta T_{TSZ}\rangle_\rho/\langle F\star \Delta T_{TSZ}\rangle_{\rho}$
where the average is on a small disc of radius $\rho$ around the cluster
center in both the unfiltered and filtered maps.
In Table~\ref{table1} we list the  
monopole  as measured within an aperture of  5\arcmin\ radius in the 
original WMAP data and in the KAB, OMCP, and MH filtered maps. The quoted
values correspond to the mean and rms dispersion of the four W DAs. 
We divide our sample according to X-ray luminosity in three independent bins
to test the correlation of our measurements with X-ray luminosity as 
in Kashlinsky et al (2010). Within 5\arcmin\ the measured $R^F$ 
is rather uncertain since the temperature anisotropy 
contains CMB residuals and noise together with TSZ. This is specially
evident for the less massive clusters (sample I).
From Table~\ref{table1}, the renormalization coefficient
is $\sim 2$ for the KAB filter,
$\sim 600$ for the MH filter whereas the OMCP filter does not dilute but boosts  
the signal (but not the S/N) by about a factor of 3. Since all clusters
in our sample have been fit to a $\beta$-model, we can use this profile
to compute numerically $R^F$. It should be computed separately for each cluster
sample and DA. Nevertheless, \textit{to avoid introducing an extra variance, we will use
the same values of $R^F$ for all cluster configurations and bands.} In this way,
we do not change the S/N ratio of the measured dipoles when expressed in $\mu$K. 
For the full sample the ratios are $R^F\simeq[1.7,0.37,530]$ for the KAB, OMCP
and MH filters, respectively. Since different bands have different normalization
coefficients and different zero monopole aperture, we concentrate our
study on the 4 DAs of the W-band. Although  
the results averaged over all WMAP bands are very similar due to the noise, by 
using only the W-band, the statistical significance of the measured
dipole is slightly decreased. For the purpose of this paper this is
not important since we are only interested in comparing the S/N ratio
of the three filters.

\begin{table}[t]
\begin{center}
\begin{tabular}{|ll|c|cccc|ccc|}
\hline
\multicolumn{2}{|c|}{Cluster Sample} & $N_{\rm cl}$ & 
\multicolumn{4}{|c|}{$\langle\Delta T\rangle$[5']/$\mu$K} & 
\multicolumn{3}{|c|}{$R^F\langle\Delta T\rangle$[30']/$\mu$K} \\
        &          &           & \multicolumn{1}{|c}{WMAP} &
\multicolumn{1}{c}{KAB} & \multicolumn{1}{c}{OMCP} & \multicolumn{1}{c|}{MH} &
KAB & OMCP & MH\\
\hline
I & $0.5<L_X<1$ & 208    & 
$-14\pm 9$ & $-4\pm 7$ & $-24\pm 26$ & $-0.03\pm 0.01$
& -1.5$\pm$0.9  &  -2.3  $\pm$1.0 & -8.3$\pm$1.9  \\

II& $1<L_X<2$& 179    & 
$-21\pm 7$ & $-10\pm 8$
& $-50\pm 28$ &$-0.02\pm 0.008$ & 2.4$\pm$0.7 & -0.8$\pm$0.9 & -1.9$\pm$1.5  \\

III& $L_X>2$ & 119    & 
$-40\pm 10$ & $-30\pm 5$
& $-138\pm 40$ &  $-0.09\pm 0.02$ & 1.0$\pm$1.5 & -4.2$\pm$1.3 & -13$\pm$2  \\
\hline
\end{tabular}
\caption{Different cluster samples considered in this work and some of their
properties. $N_{cl}$ is the number of clusters in each subsample and
X-ray luminosities are given in units of $10^{44}$erg s$^{-1}$.}
\label{table1}
\end{center}
\end{table}

\begin{figure}
\plotone{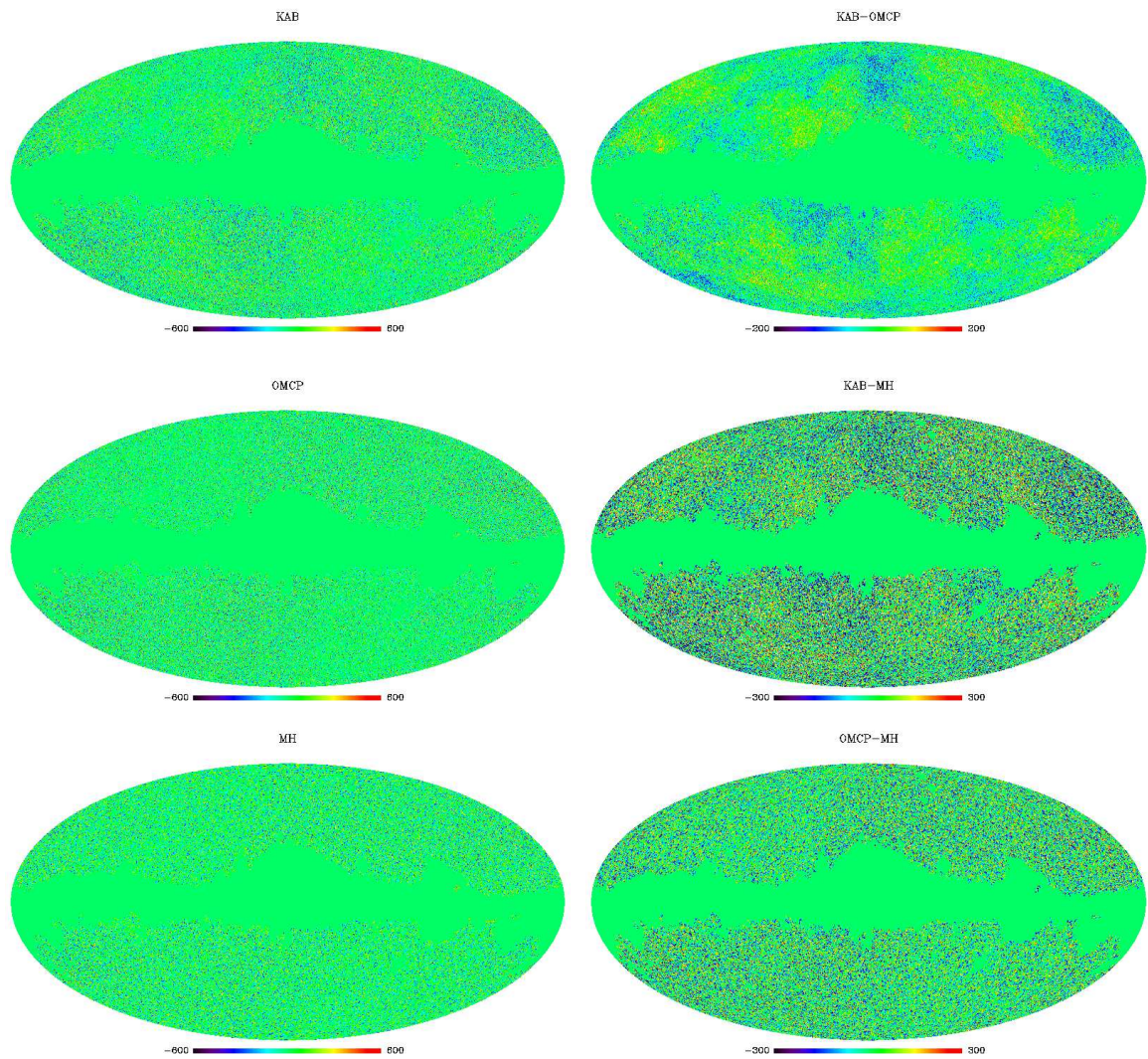}
\caption{\small Filtered maps in Healpix format.
Left column and from top to bottom: W1 DA map filtered using the
KAB, OMCP and MH filters. Right column: differences between the
various filtered maps.}
\label{fig:fig4}
\end{figure}

Our normalization is useful for comparing 
the results of different filters since all filtered maps have similar power. 
It enables us to subtract maps from each other to compare their different structure. 
In Fig.~\ref{fig:fig4} (left column) we show the KAB, OMCP, and MH filtered maps of the 
W1 DA. The KAB map exhibits more structure on large scales since our filter preserves 
power below $\ell=300$. The OMCP filter erases all power at $\ell\la 200-300$, and 
hence the resulting filtered maps show no large-scale features. The MH filter, 
finally, preserves power over a wide range of scales ($50\la \ell\la 600$) 
but not on the largest scales. Consequently, the MH filtered map shows 
pronounced structure only on intermediate scales. The differences
in the impact the filters have on the CMB signal become  
clearer when the filtered maps are subtracted from each other.
The subtracted maps are shown in the right column of Fig.~\ref{fig:fig4}. 
As expected, the difference between the KAB and OMCP maps is most prominent on 
large angular scales. The differences between the OMCP and MH filtered maps 
are dominated by features on intermediate scales to which the MH filter is 
sensitive, whereas the OMCP filter is not. The difference between the KAB 
and MH maps, finally, exhibits a more complex pattern that combines the 
effects of the KAB filter's sensitivity to very low multipoles (large scales) 
and the MH filter's emphasis on intermediate scales (cf.\ Fig.~\ref{fig:filters}).

\subsection{Cluster-by-cluster comparison of the filtering schemes}

Any meaningful filter should significantly remove the primary CMB component 
or else it would not
be able to isolate the much smaller KSZ dipole term lurking beneath it.
The TSZ component also needs to be lowered. Filters redistribute differently
the power of the intrinsic CMB anisotropies, TSZ and KSZ.
Before discussing the efficiency of each filter we investigate empirically 
how each of the filtering schemes transfers the TSZ and primary CMB components 
with CMB temperature, averaged over apertures of size $\rho=$10\arcmin\ 20\arcmin\ 
and 30\arcmin, at each cluster location $\hat{n}_j$ on the sky, $\langle\Delta T
(\hat{n}_j)\rangle_\rho$.
The all-sky mean of all $\langle \Delta T\rangle$ is the monopole that we 
use to define the zero-monopole aperture. We will illustrate our 
findings with the W1 DA data which has the best angular
resolution. By choosing the data at the same DA, we make sure that all
three datasets have the same instrument noise realization on a per pixel basis.

\begin{figure}
\plotone{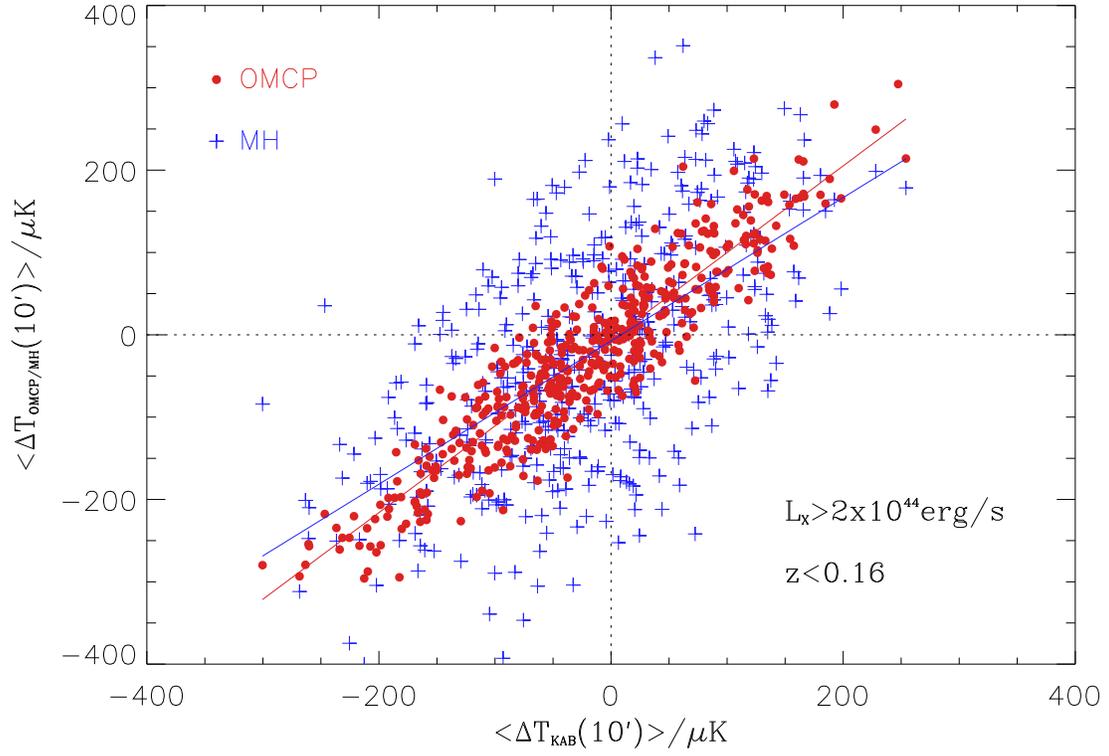}
\caption{\small Mean temperature anisotropy on a disc of radius
10' at each cluster location for clusters of sample III.
Red circles correspond to OMCP vs KAB filter
and blue crosses to MH vs KAB. Straight lines were obtained by
minimum square regression.}
\label{fig:fig5}
\end{figure}

In Fig.~\ref{fig:fig5} we plot the mean temperature anisotropy in the 
central parts of all the clusters in our sample III of Table~\ref{table1}, 
corresponding to the aperture radius $\rho=10$\arcmin, comparing 
the OMCP- and MH-filtered maps with the KAB-filtered data.
The figure shows that on average the same CMB components, i.e. the primary CMB and TSZ
that are present in the central parts of clusters in the KAB-filtered
maps, are also present in the other two filtered maps. 
At the same time, there are also notable differences which manifest themselves 
in the dispersion shown in the plot. The dispersion of the MH- vs KAB-filtered
data is much larger than in the OMCP vs the KAB case. The correlation coefficient
between KAB and OMCP is 0.94 dropping to only 0.62 for KAB and MH.
This comparison shows empirically that the filters  
redistribute the profile (and each of the components) very differently. 

The primary CMB and TSZ components present at $\rho=10$\arcmin\
need to be removed if there is any hope of detecting the KSZ-like dipole.
Fig.~\ref{fig:fig6} compares the mean temperature over $\rho=20$\arcmin\ and 30\arcmin\
with that at $\rho=10$\arcmin\ for the W1 DA in two redshift bins
of roughly the same number of clusters. 
Fig.~\ref{fig:fig6}a-c correspond to our samples II and III,
and Fig.~\ref{fig:fig6}d-f to clusters in our sample not considered
in this study.  The plots 
illustrate empirically that KAB removes the intervening components 
better than the other two filters. 
For the KAB filter, there is little to no correlations at
these larger radii with the central part, consistent with
this filter having removed the components present at 10\arcmin, {\it 
the main obstacle to measuring the KSZ-like dipole}. 
The OMCP filter (middle panels) shows the presence of these components at  the
$\rho >20$\arcmin\ aperture radii and thus it is not as efficient 
as the KAB filter. The MH filter performance is much more dismal; it shows
an almost perfect correlation between the larger apertures and that at 10\arcmin,
implying that the same components that have prevented the measurement of
the KSZ signal in cluster centers remain, although redistributed, as far out
as $\rho>30$\arcmin. 

\begin{table}
\begin{center}
\begin{tabular}{|l|c|ccc|ccc|}
\hline
Filter & $\sigma_0/\mu$K & $D_X/\mu$K & $D_Y/\mu$K & $D_Z/\mu$K 
& $D/\mu$K & $l$/1\degr & $b$/1\degr \\
\hline
KAB & 30& $-3.0\pm 2.7 $&$ -5.5\pm 2.4$ &$ 1.8\pm 2.0 $&
$6.5\pm 2.3$ & $241\pm 27$  & $16\pm 18$ \\
OMCP & 23& $-0.7\pm 2.0 $&$ -2.0\pm 1.8$ &$ 1.2\pm 1.5 $&
$2.4\pm 1.4$ & $251 \pm 67$ & $30\pm 28$ \\
MH &67&$-2.6\pm 5.7$ & $-7.0\pm 5.3$ & $4.1\pm 4.4$ &
$8.5\pm 4.4$ & $249\pm 57$ & $29\pm 26$ \\
\hline
\end{tabular}
\end{center}
\caption{Dipole components, moduli and directions for the three filters.}
\label{table2}
\end{table}

\subsection{Dipoles.}\label{sec:dipoles}

To compare the performance of the three filters, we computed dipoles for 
each filter and for the three cluster samples of Table~\ref{table1}. We used
the following apertures of radius $\rho=[5,10,20,30,40,50,60]$ arcmin\ in order 
to select a suitable zero-monopole aperture for each filter. As indicated,
all data are renormalized by multiplying 
the filtered KAB, OMCP and MH maps by the coefficients $R^F=[1.7,0.37,530]$,
respectively. With this normalization, the monopole evaluated on
apertures of 5\arcmin\ would have the same amplitude in the three filtered
maps. Ideally, this would also be true for the dipole. By using a single
coefficient, the S/N of the dipole measurement in each cluster
configuration remains unaltered.  

In Table~\ref{table1} we also list the value
of the monopole residual at 30\arcmin\ in the renormalized maps.
The KAB zero monopole aperture occurs at 
$\le 30$\arcmin\ radius, for OMCP is at $\sim$30\arcmin-40\arcmin\ and for MH is at 
$\ge$50\arcmin; Table~\ref{table1} shows that at 30\arcmin\ 
the MH filter still contains non-zero monopole residuals, in 
agreement with Fig.~\ref{fig:fig6}.
In Figs.~\ref{fig:dipole}a-c we plot the three components (X,Y,Z) 
of the dipoles measured with an aperture of 30\arcmin\
versus the TSZ monopole at 10\arcmin\ radius. In Fig.~\ref{fig:dipole}d we plot
the dipole moduli. The three cluster subsamples of Table~\ref{table1} are 
represented with the same symbol. They can be easily distinguished
since the average TSZ monopole increases with increasing X-ray luminosity.
Notice that each cluster subsample has similar monopoles at 10\arcmin\ for the
three filters as expected since our normalization was constructed to produce 
the same TSZ amplitude in the central parts of clusters at 5\arcmin.

The errors on the measured dipoles $\sigma_i$, $i=(X,Y,Z)$, were obtained 
from the dispersion of dipoles measured in random cluster positions in the sky. 
For each component, the error scales approximately as (Atrio-Barandela et al 2010)
\begin{equation}
\sigma_i^2=\frac{(\sigma^{\rm res\;CMB}_i)^2}{N_{\rm cl}}
+\frac{(\sigma^{\rm noise})^2}{N_{\rm pix}N_{\rm DA}}
\label{eq:sigmas}
\end{equation}
where $\sigma^{\rm res\;CMB}_i$ is the rms dispersion of the CMB residual
in the filtered map, $\sigma^{\rm noise}$ the dispersion of the noise,
$N_{\rm pix}$ the number of pixels occupied by the cluster sample and
$N_{\rm DA}$ the number of DAs used in the analysis. We considered four cluster
samples with different numbers of clusters (100, 200, 400 and 800 clusters). For 
each sample, we generated 4,000 random cluster distributions. If clusters sampled 
the sky homogeneously then the error on the monopole, $\sigma_0$, and in each dipole 
component would be related by $\sigma_i=\sqrt{3}\sigma_0$. 
For an inhomogeneous cluster distribution the relation is given by
$\sigma_i\simeq\sigma_0/\langle n_i^2\rangle^{1/2}$ where 
$\langle n_i^2\rangle$ is the average of the direction cosines of the clusters
in the sample (Atrio-Barandela et al. 2010, Atrio-Barandela 2013). 
Since the Galaxy removes a large fraction of the sky, preferentially
in the X direction, errors are larger for $D_X$ than
for $D_Y$ or $D_Z$. We verified that this was the case for the three
filters, being 10-18\% and 2-3\% larger 
with respect to a homogeneous distribution of clusters
for X and Y, respectively, and a 15\% smaller for Z. Since the
errors are dominated by the residual CMB, they 
scale as $N_{cl}^{1/2}$, as dictated by eq.~(\ref{eq:sigmas}). For each
cluster sample, the error on each dipole component is scaled 
by the number of clusters from the error obtained on the simulations.
The errors on the modulus of Fig.~\ref{fig:dipole}d were obtained from 
the rms dispersion of 10,000 dipoles. Those dipoles were generated 
by adding to the each measured dipole component a random value drawn
from a gaussian distribution with zero mean and dispersion $\sigma_i$.

The results from the three filters show a reassuring degree of consistency 
regarding the value of the Y-component of the dipole. This component is always 
negative for all filters and all cluster configurations, in agreement with 
the DF results, which combine a pronounced negative Y-component with a positive 
Z-component of lower amplitude and an X-component that is consistent with 
zero within the measurement uncertainties. The KAB filter (blue
filled circles) shows a very clear correlation of the Y dipole component 
with the monopole at 10\arcmin\ (Fig.~\ref{fig:dipole}b) and so does the overall
dipole amplitude (Fig.~\ref{fig:dipole}d). This correlation is found at 
30\arcmin, when the TSZ monopole is zero. Then, there can not be a TSZ dipole 
contribution due to an inhomogeneous cluster distribution. 
It is instructive to revisit Fig.~\ref{fig:fig6}a: cluster by cluster,
the average temperature anisotropy on a disc of 10\arcmin\ is uncorrelated with
the same average over a disc of 30\arcmin. 

Fig.~\ref{fig:dipole}a also shows a strong correlation for X-component 
in the MH filter (green triangles). However, in this case we expect 
this correlation to be due to the TSZ effect. As our error analysis show, 
the X-component of the dipole is the worst measured since the Galaxy is most 
predominant in this direction. Then, one has to expect an important TSZ 
dipole due to the inhomogeneous distribution of clusters in the sky
in agreement with the fact that for the brightest clusters, sample III, 
the residual TSZ monopole at 30\arcmin\ is 
$\langle\Delta T(30')\rangle=-13\pm 2\mu$K, large in comparison
with the value of the other two filters (see Table~\ref{table1}). 
It is also in agreement with 
the result of Fig.~\ref{fig:fig6}c that shows that the 
average temperature anisotropy at cluster locations within discs of
radius 30\arcmin\ (TSZ monopole) is strongly correlated with the same 
average at 10\arcmin\ and, consequently, so it would be the TSZ dipole at 30\arcmin.
We verified that the measured dipoles are in fact very small at the 50\arcmin\ aperture,
the zero monopole aperture for this filter.

With respect to the OMCP filter (open red squares) the dipole is always
small and shows no correlation with the monopole at 10\arcmin. 
The low dipole amplitudes were to be expected since 
this filter cuts off all the power at multipoles $\ell\le 300$ 
(compare Fig.~\ref{fig:signal}b and Fig.~\ref{fig:filters}b) and therefore
is insensitive to half the dipole signal. At this filter zero 
monopole aperture, $\sim 40$\arcmin\ the results were
not very different, showing no significant dipole either.

\subsection{Consistency of the measured dipoles.}

Since the three cluster samples of Table~\ref{table1} are independent
the motion of all the clusters can be obtained by averaging the
measured dipoles $D_{i,s}$ in each sample $s=(I,II,III)$ weighted by their respective 
errors $\sigma_{i,s}$
\begin{equation}
D_i=(\sum_s D_{i,s}/\sigma_{i,s}^2)/(\sum_s 1/\sigma_{i,s}^2)
\end{equation}
In Table~\ref{table2} we give the resulting dipoles for the three filters.
We also give $\sigma_0$, the rms dispersion of the monopole.
Consistently, the error bars on the dipoles are 
$\sigma_{(X,Y,Z)}\simeq\sigma_0\sqrt{3/N_{cl}}$, being larger for the X
and smaller for the Z components, as we found with the simulated dipoles
(see Sec~\ref{sec:dipoles}). In the table we also give the dipole
amplitude and direction for the three filters. The errors were computed
from the dispersion of 10,000 dipole moduli and directions as
in Fig~\ref{fig:dipole}d. Notice the remarkable agreement between
the direction of the three filters. The angular separation 
between the central value of the direction of the KAB filter and that of OMCP and MH is
about 16\degr. More remarkable is the close alignment between the
direction of the OMCP and MH dipoles. However, the errors on the direction
are large and we can not derive a final conclusion of this coincidence 
since the MH dipoles have probably an important contamination from
the TSZ monopole, as discussed in Sec~\ref{sec:dipoles}.

The results of Table~\ref{table2} are crucial to understand the performance of 
each filter. Of the three filters, the KAB filter
has the largest S/N. By effectively removing all multipoles with $\ell\le 300$,
OMCP filtered maps contain a small fraction of CMB residuals and
give the smallest error bars. Unfortunately, the filter is also
very effective in removing the dipole signal, degrading the overall S/N.
Thus, our results agree with the original OMCP analysis since 
velocities above 4,000km/s were required on samples of $\sim 700$ clusters
for the dipole to be detectable. By comparison, the MH filter measures
dipoles of larger amplitude but the error bars are also large. Only the
KAB filter provides statistically significant result. For this cluster
sample our measurement is significant at the
93\% confidence level, smaller than in Kashlinsky et al (2010)
(see Atrio-Barandela et al, 2013, for a full discussion)
while the results of the OMCP and MH have a significance below 60\%,
reflecting the poor performance of these two filters.

\subsection{From temperature to velocity.}

To compare properly the results of the different filters, dipoles
need to be expressed in velocity and not in temperature units.
In Kashlinsky et al (2009) we denoted the calibration coefficient as:
$C_{1,100}=(D/V_{\rm Bulk})=(T_0/c)\langle F\star\tau\rangle$, with
$D$ the dipole in $\mu$K and $\langle F\star\tau\rangle$ 
the \textit{filtered} cluster profile averaged 
solid angle subtended by the cluster population. 
For the correct profile, $V_{\rm Bulk}=D/C_{1,100}$ would be
independent of the aperture used to measure the dipole.
To account for the dilution effect introduced by the filter 
our filtered maps have been renormalized by $R^F$ (see Sec~3.1).
In the renormalized maps, $C_{1,100}=(T_0/c)\langle\tau\rangle$.
Using the $\beta$ model data, we can compute $\langle\tau\rangle$
for our cluster sample for different apertures. We find that
for $\rho=5,10,15$\arcmin, the coefficient averaged over the
four W DAs is $C_{1,100}=(1.36\pm0.08,\,1.07\pm0.05,\,
0.86\pm0.03)[\mu$K$/100kms^{-1}$], respectively. Since
Fig.~\ref{fig:fig5} shows that at the 10\arcmin\ aperture
the three renormalized filtered maps have similar monopoles at 
the cluster locations, we can assume that $\langle\tau\rangle$
is also similar for the three filters out to that aperture.
Then, by taking the conversion factor for 10\arcmin\ we can translate
the dipole into a velocity
\begin{equation}
V_{\rm Bulk}^{\rm KAB}=607\pm 215\,{\rm km\,s}^{-1}\,,\quad
V_{\rm Bulk}^{\rm OMCP}=224\pm 131\,{\rm km\,s}^{-1}\,,\quad
V_{\rm Bulk}^{\rm MH}=794\pm 411\,{\rm km\,s}^{-1}\,.\quad
\label{eq:velocities}
\end{equation}
With this normalization, the dipole amplitude in \kms\ 
is lower than the value of $\sim 1000$ \kms\ found previously
(see Kashlinsky et al 2009, 2011 for a full discussion on our previous method).
Nevertheless, our calibration of the flow is still not exact since the $\beta$ model
is not an accurate description of cluster profiles outside the X-ray emitting inner part.
The dipole is measured at 30\arcmin\ while $C_{1,100}$ is computed 
at 10\arcmin. If $C_{1,100}$ is smaller at 30\arcmin, as suggested
by the calculation at 15\arcmin, then the corresponding velocities
will be larger. For instance, if $C_{1,100}\sim 0.55 [\mu$K/$100 kms^{-1}]$
then the velocities will be those of our previous work.

Even if our calibration of the flow is still uncertain
due to lack of proper data, it does not affect the statistical
significance of the dipole measured with any of the three filters.
The KAB filter does not only measures a dipole that correlates
with cluster X-ray luminosity; it provides the largest signal to noise ratio
of the three filters discussed here. Nevertheless, the discrepancies
between the three filters could be real if 
the three filters are probing the flow at different scales. 
For instance, since the OMCP filter is sensitive primarily to scales corresponding 
to $\ell\ge 300$, it gives all the weight to unresolved clusters which are, 
on average, farther away. By contrast, the KAB dipole includes 
contributions from nearby systems, owing to the filter's greater 
sensitivity to low multipoles. While some degree of discrepancy between 
filters could simply reflect spatial variations in the
properties of the flow, the fact is that none of the filters but KAB
has been shown to be limited by cosmic variance, has
provided a statistically significant measurement or has
shown a correlation with X-ray luminosity at the zero monopole aperture.
Therefore, the results obtained using the OMCP or MH filters can
not be interpreted as ruling out any possible large scale flow detected
by the KAB filter.

\section{Summary and Conclusions.}\label{sec:summary}
\begin{figure}
\plotone{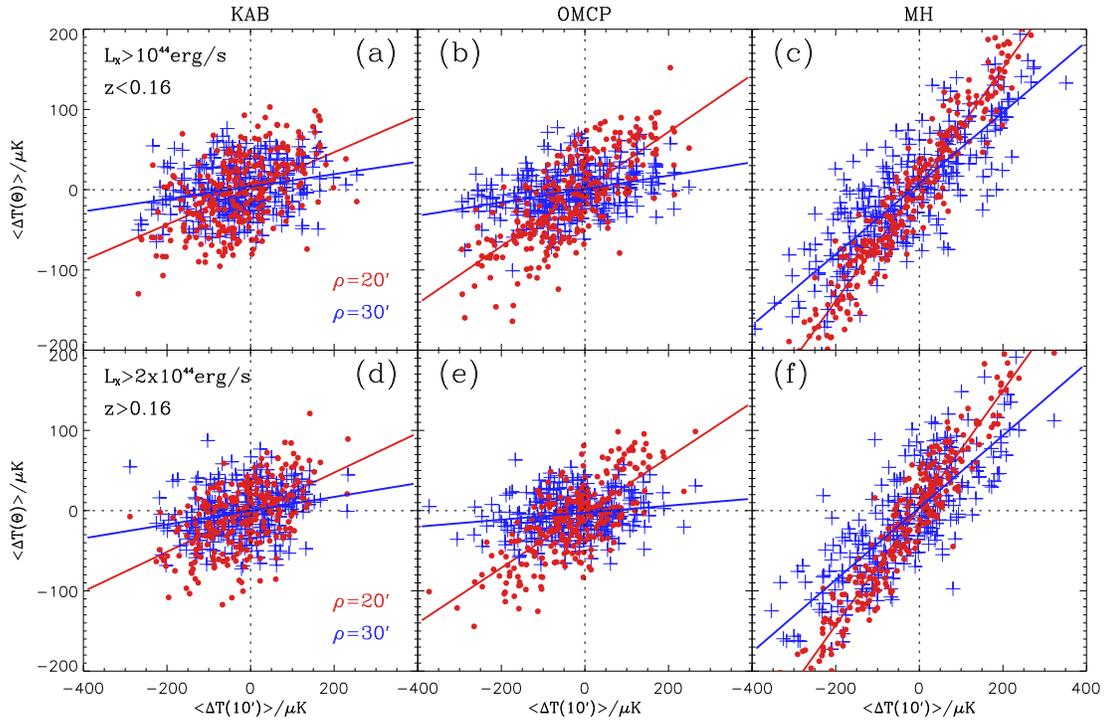}
\caption{\small 
Mean temperature anisotropy on the filtered W1 DA,
averaged over discs of radius $\rho=20$\arcmin\ and 30\arcmin\ versus
10\arcmin, for the three filters and for two cluster configurations.
Red dots and blue crosses correspond to 20\arcmin and 30\arcmin, respectively.
The solid line is the minimum square regression fit.}
\label{fig:fig6}
\end{figure}

\begin{figure}
\plotone{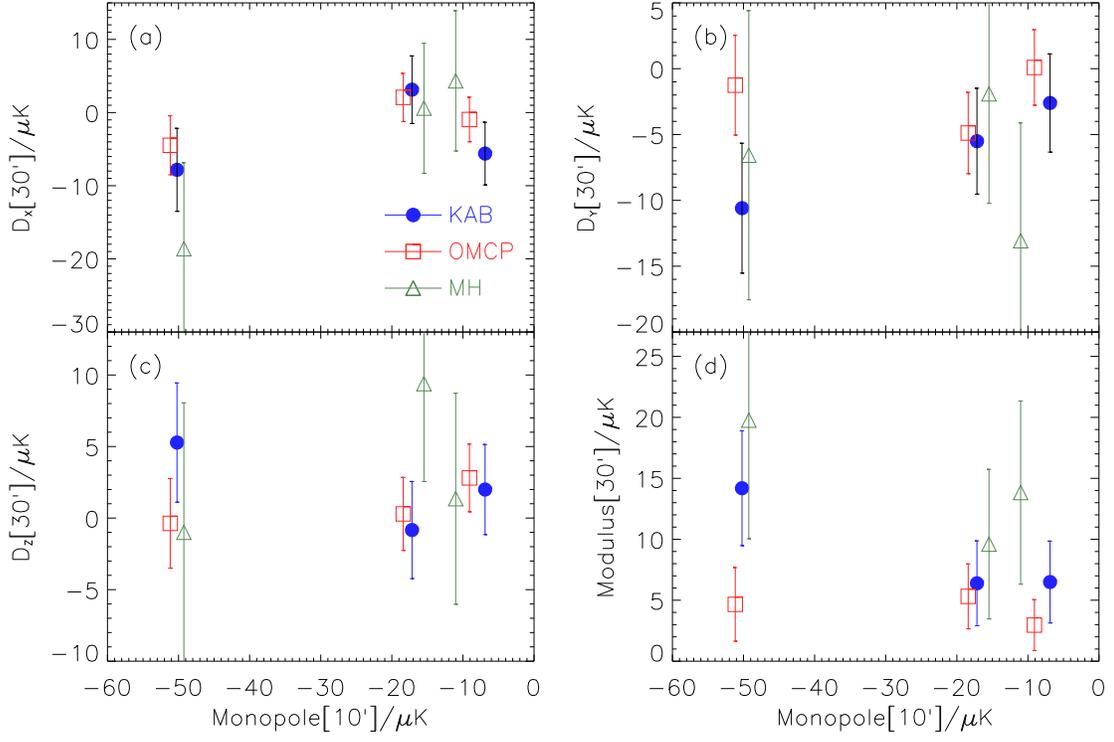}
\caption{\small 
X, Y, and Z components of the dipole and modulus of the dipole 
evaluated within 30\arcmin\ for the  KAB (filled circles), OMCP 
(open squares) and MH (triangles) filters, respectively, versus the monopole at 10\arcmin\
aperture for the three samples given in Table~\ref{table1}. Error bars are
obtained from 4,000 simulations of clusters templates randomly placed on
the filtered maps.
}
\label{fig:dipole}
\end{figure}

We can now proceed with the summary of the appropriateness of the three
filtering schemes in uncovering the KSZ-like dipole signal in the KAB method.
\begin{itemize}
\item
{\bf KAB filter}. This filter is the most appropriate of the three
as has already been explained in the original KAB paper. It removes
the primary CMB down to the fundamental limit imposed by the cosmic
variance as demonstrated analytically and numerically in
Atrio-Barandela et al (2010). It has the smallest zero monopole
aperture of the three filters and, as Fig.~\ref{fig:fig6} shows, at 
the final - zero monopole -
aperture the CMB does not correlate in any meaningful manner with
that at the central parts. This therefore allows to look for the KSZ dipole
component after the TSZ has been made vanishingly small because
of the cluster X-ray temperature decrease toward the outer parts
(Kashlinsky et al 2009). The dipole uncovered at cluster positions at such
aperture contains a negligible TSZ component, which dominates the
central parts at $\rho=10$\arcmin. The filter is sensitive to
a KSZ dipole spread over the whole $\ell$ space and 
shows the largest S/N of all the filters anlyzed here.
\item
{\bf OMCP filter} is clearly not as efficient as the KAB. By cutting 
power at $\ell\le 300$ it reduces the residual CMB that dominates
the error bar but also decreases the signal, reducing the overall S/N
of the filter. This point was already
demonstrated implicitly in Fig.~13 of OMCP which shows
that no bulk flow can be reliably measured with this filter at velocities
$< 4,000-10,000$ km/sec; we find it strange that, given those
numbers, claims are occasionally still made about the alleged efficiency of
this filter. Fig.~\ref{fig:fig6} and Table~\ref{table2} demonstrate explicitly that if
one uses this filter, particular care must be made to work at the
appropriately large apertures of at least $>30$\arcmin\ in radius.
Nevertheless, as the filter is only sensitive to a rather small
range of multipoles ($\ell\ge 300$), even at these large apertures the
signal has been erased as to make the dipole undetectable. 
\item {\bf MH filter}. This filter is the least useful for the application
to the KAB method. Even at apertures as large as $>30$\arcmin\ it
still leaves primary CMB and TSZ components, as demonstrated by the
clear correlation in Fig.~\ref{fig:fig6}c, that confuse any
determination of the KSZ-like dipole signal. Furthermore,
the filter does detect a dipole at the 30\arcmin\ aperture, but due
to the large monopole residuals, there exists important contributions
due to TSZ dipoles. The measured dipoles have the largest error bars
and, like in the OMCP filter, the results are not statistically significant.
\end{itemize}

To conclude, we have compared the dipoles measured at the locations of X-ray 
selected galaxy clusters using the filters
proposed by OMCP and MH with the results obtained with the KAB filter.
{\it We find that all filters recover dipoles; the central
values of the directions are aligned but the KAB dipole is the most efficient since
it has the largest statistical significance.} The discrepancy between these
findings of ours and the results published by OMCP and MH
is due to the latter teams failing to chose the right aperture for their
measurement. In addition, their simulations and calibrations assume 
the same profile for the electron density and electron pressure, hypothesis
that is bound to degrade their filter performances if real clusters
are not described by their profiles.

In comparison with the other filters,
the KAB filter makes no assumptions about cluster profiles or extents, 
and by design removes the intrinsic CMB, the largest contaminant 
(see eq.~\ref{eq:dipole}) down to the limits imposed by cosmic variance. 
The impact of all uncertainties associated with our limited knowledge 
of the cluster profile and extent is limited to a single normalization 
coefficient $C_{1,100}$ that does not
affect the filter definition. In addition, the zero-monopole 
aperture for the KAB filter has a radius of $\la 30$\arcmin, smaller than 
what is required by the other two filters. Finally, the dipole signal recovered 
with the KAB filter is strongly correlated with the central TSZ monopole 
(or the equivalent X-ray luminosity, Kashlinsky et al 2010),  
a clear indication that the dipole is not produced by some random
contribution of the residual CMB. These three aspects make the KAB filter
the least affected by possible systematics among the 
filters suggested to date. Alternative filtering schemes 
can possibly be designed, but their design must ensure that
they increase the S/N of the measurement, not decrease it.
Finally, filters redistribute the information from real to filtered
space by weighing angular scales differently. Therefore, they are probing
the velocity field on different scales and the discrepancies on
modulus and direction could reflect the spatial variations of the flow.
To demonstrate that filters probe the same or different flows would 
require accurate cluster profiles and it goes beyond the scope of the present work.

Recently, and well after this paper had been submitted to peer-review, 
the Planck Collaboration carried out the analysis of
the filters described here using the recently released Planck
data (Planck Intermediate Result XIII, Planck Collaboration, 2013). 
The KSZ team analyzed peculiar velocities using the Internal Linear Combination 
(ILC) map, constructed by combining data at different frequencies with varying
weights in $\ell$-space in order to reduce the foreground contribution. 
The results found using the KAB
filter were in agreement with ours but the Planck Collaboration 
assigned larger statistical uncertainties, reducing their statistical significance. 
The flaws of their analysis, that lead to an overestimation of their
errors, are discussed in Atrio-Barandela (2013). Also, the ILC data
is not the best suited data for this analysis. Since the weights of
different frequencies vary with $\ell$, the power of the SZ signal is scrambled
across different multipoles and makes it very difficult to
establish how filters that operate on $\ell$-space transfer the cluster signal.
Peculiar velocity measurements will be better studied using Planck frequency data.
Frequency information will be critical to establish the origin, nature, and scale 
of the dipole measured in the WMAP data. Cluster profiles can
be measured individually to determine the effect of filtering on each of the clusters;
the frequency coverage would be crucial to remove any TSZ dipole component and,
specifically, the 217 GHz channel would be important to test for systematics.

\acknowledgements
FAB acknowledges financial support from the Spanish
Ministerio de Educaci\'on y Ciencia (grants FIS2009-07238, FIS2012-30926
and CSD 2007-00050). HE gratefully acknowledges funding 
provided by NASA grant NNX10AJ69G.

\pagestyle{plain}

\end{document}